\documentclass[prc,preprint]{revtex4}
\usepackage{epsfig}
\begin{document}
\flushbottom

\title{Impact parameter dependence of heavy ion $e^+ e^-$ pair production to
all orders in $Z \alpha$}
\author{A. J. Baltz}
\affiliation{Physics Department,
Brookhaven National Laboratory,
Upton, New York 11973}
\date{September 25, 2006}

\begin{abstract}
The heavy ion probability for continuum $e^+ e^-$ pair production has been
calculated to all orders in $Z \alpha$ as a function of impact parameter. 
The formula
resulting from an exact solution of the semiclassical Dirac equation in the
ultrarelativistic limit is evaluated numerically.  In a calculation of
$\gamma = 100$ colliding Au ions the probability of $e^+ e^-$ pair production
is reduced from the perturbation theory result throughout the impact parameter
range.
\\
{\bf PACS: { 25.75.-q, 34.90.+q}}
\end{abstract}
\maketitle
\section{Introduction}

The impact parameter dependence of heavy ion $e^+ e^-$ pair production to all
orders in $Z \alpha$ is of direct relevance to experiments carried out at the
BNL Relativistic Heavy Ion Collider (RHIC) and possible at the CERN Large
Hadron collider (LHC).  In the usual method of scattering theory, the impact
parameter is formally integrated over to turn the momentum transfer of the
amplitude into that of the cross section via a delta function\cite{pes}.
However a standard method of triggering measured events at RHIC and LHC is with
zero degree calorimeters, devices that respond to neutrons from Coulomb
dissociation of the beam ions\cite{bcw}.
The physical process of a measured pair production event then also includes
simultaneously the Coulomb dissociation of the ion nucleus.  Assuming
independence of the two processes (pair production and Coulomb dissociation)
the probability of an event is then the product of the probability of pair
production times the probability of Coulomb dissociation.
Predicted cross
sections are then appropriately constructed from the integral of the product
of the pair production probability times the dissociation probability.
I will confine myself in this paper to evaluation of the impact parameter
dependent probability for heavy ion production of $e^+ e^-$ pairs.  Combining
the probability of the particular $e^+ e^-$ phase space acceptance of
physical detectors with probability of Coulomb dissociation will be left
for future treatment.

The impact parameter dependent calculation of heavy ion $e^+ e^-$ pair
production to all orders in $Z \alpha$ described in this paper is based
on techniques previously developed.  That development
began with the realization that in an appropriate gauge\cite{brbw},
the electromagnetic field of a relativistic heavy ion is to a very good
approximation a delta function in the direction of motion of the heavy ion
times the two dimensional solution of Maxwell's equations in the transverse
direction\cite{ajb1}, and that an exact solution of the
appropriate Dirac equation could be obtained for excitation of bound-electron
positron pairs exhibiting a reduction from perturbation theory of a little less
than 10\% for Au + Au at RHIC \cite{ajb2}.

An analytical solution of the Dirac equation was subsequently
obtained\cite{sw1,bmc,sw2} for the analogous case of continuum
$e^+ e^-$ pair production induced by the corresponding countermoving delta
function potentials produced by heavy ions in an ultrarelativistic collider
such as RHIC. It was noted at the time\cite{bmc,sw2} that the
exact solution appeared to agree with the perturbation theory result.

Several authors then argued \cite{serb,lm1,lm2} that a correct
regularization of the exact Dirac equation amplitude should lead to
a reduction of the total cross section for pair production from
perturbation theory, the so called Coulomb corrections.  The first
analysis was done in a Weizsacker-Williams approximation\cite{serb}.
Subsequently, Lee and Milstein computed\cite{lm1,lm2} the total cross
section for $e^+ e^-$ pairs using approximations to the exact amplitude
that led to a higher order correction to the well known Landau-Lifshitz
expression\cite{ll}.  One might also have noted that it is well known that
photoproduction of $e^+ e^-$ pairs on a heavy target shows a negative
(Coulomb) correction proportional to $Z^2$ that is well described by the
Bethe-Maximon theory\cite{bem}.  In fact some years ago Bertulanni and
Baur\cite{bb} applied the structure of the Bethe-Maximon theory to obtain an
impact dependent probability for heavy ion reactions with a negative Coulomb
correction.  I will comment on their results and compare them with the results
of the present work in Section V.

In a previous paper I have tried to
explicate the Lee and Milstein approximate results and argued their
qualitative correctness\cite{ajb3}.  Subsequently I undertook the
full numerical calculation of electromagnetically induced ultrarelativistic
heavy ion electron-positron pair production\cite{ajb4}.
That evaluation of the ``exact'' semi-classical
total cross section for $e^+ e^-$ production with gold or lead ions
showed reductions from perturbation theory of 28\% for the CERN Super Proton
Synchrotron (SPS) case, 17\% for RHIC,
and 11\% for LHC.
For large Z no final momentum region was found in which there was
no reduction or an insignificant reduction of the exact cross section
from the perturbative cross section.

In the present paper I reformulate the approach used for my previous
cross section calculations in order to calculate corresponding
impact parameter dependent probabilities.

\section{Impact parameter dependent probabilities}

For production of continuum pairs in an ultrarelativistic heavy ion reaction
one may work in a frame of two countermoving heavy ions with the same
relativistic $\gamma$, and the electromagnetic interaction arising from them
goes to the limit of two $\delta$ function potentials
\begin{equation}
V(\mbox{\boldmath $ \rho$},z,t)
=\delta(z - t) (1-\alpha_z) \Lambda^-(\mbox{\boldmath $ \rho$})
+\delta(z + t) (1+\alpha_z) \Lambda^+(\mbox{\boldmath $ \rho$}) 
\end{equation}
where
\begin{equation}
\Lambda^{\pm}(\mbox{\boldmath $ \rho$}) = - Z \alpha 
\ln {(\mbox{\boldmath $ \rho$} \pm {\bf b}/2)^2 \over (b/2)^2}.
\end{equation}

I calculate the number weighted probability (or number operator) $<N(b)>$ for
producing $e^+ e^-$ pairs at some impact parameter $b$
\begin{equation}
<N(b)> = \sum_{n=1}^{\infty} n P_n(b) = \int {m^2 d^3 p\, d^3 q 
\over (2 \pi)^6 \epsilon_p \epsilon_q } \vert M(p,q) \vert^2
\end{equation}
with the previously derived exact semiclassical amplitude for
electron-positron pair production\cite{sw1,bmc,sw2} written in the
notation of Lee and Milstein\cite{lm1}
\begin{equation}
M(p,q) = \int { d^2 k \over (2 \pi)^2 } \exp [i\, {\bf k} \cdot {\bf b}]
{\cal M}({\bf k}) F_B({\bf k})
F_A({\bf q_{\perp} + p_{\perp} - k}),
\end{equation}
where $p$ and $q$ are the four-momenta of the produced electron and positron
respectively, $p_{\pm}=p_0 \pm p_z$, $q_{\pm}=q_0 \pm p_z$,
$\gamma_{\pm} = \gamma_0 \pm \gamma_z$,
$\mbox{\boldmath $\alpha$}=\gamma_0 \mbox{\boldmath $\gamma$}$,
${\bf k}$ is an intermediate transverse momentum transfer from the ion to be
integrated over,
\begin{eqnarray}
{\cal M}({\bf k}) &=& \bar{u}(p) {\mbox{\boldmath $\alpha$} \cdot
({\bf k -p_{\perp}})
+ \gamma_0 m  \over  -p_+ q_- -({\bf k - p_{\perp}})^2 - m^2 + i \epsilon}
\gamma_- u(-q)
 \nonumber \\
&\ & +  \bar{u}(p) {- \mbox{\boldmath $\alpha$} \cdot ({\bf k -q_{\perp}})
+ \gamma_0 m  \over  -p_-q_+ -({\bf k - q_{\perp}})^2 - m^2 + i \epsilon}
\gamma_+ u(-q),
\end{eqnarray}
and the effect of the potential Eq. (1-2) is contained in integrals, 
$F_B$ and $F_A$, over the transverse spatial coordinates\cite{sw1,bmc,sw2},
\begin{equation}
F({\bf k}) = \int d^2 \rho\, 
\exp [-i\, {\bf k} \cdot \mbox{\boldmath $ \rho$}] 
\{ \exp [-i Z \alpha \ln {\rho}] - 1 \} .
\end{equation}
$F({\bf k})$ has to be regularized or cut off at large $\rho$.
How it is regularized
is the key to understanding Coulomb corrections as I will review in
Section III.

In Section III I will show how a properly regularized transverse potential
in Eq. (6) necessitates a numerical integration over
$\mbox{\boldmath $ \rho$}$ to obtain $F({\bf k})$.  With the resultant
$F({\bf k})$
the integration over ${\bf k}$ in Eq. (4) is then not amenable to the usual
Feynman integral techniques and must be carried out numerically.  For
numerical convenience I have
chosen to express ${\bf k}$ in Cartesian co-ordinates
and rewrite Eq. (5)
\begin{eqnarray}
{\cal M}({\bf k}) &=& \bar{u}(p) {\alpha_x k_x + \alpha_y k_y
- \mbox{\boldmath $\alpha$} \cdot {\bf p_{\perp}}
+ \gamma_0 m  \over  -p_+ q_- -({\bf k - p_{\perp}})^2 - m^2 + i \epsilon}
\gamma_- u(-q)
 \nonumber \\
&\ & +  \bar{u}(p) {-\alpha_x k_x - \alpha_y k_y
+ \mbox{\boldmath $\alpha$} \cdot {\bf q_{\perp}}
+ \gamma_0 m  \over  -p_-q_+ -({\bf k - q_{\perp}})^2 - m^2 + i \epsilon}
\gamma_+ u(-q).
\end{eqnarray}

The expression for the amplitude $M(p,q)$ Eq.(4) then becomes
\begin{eqnarray}
M(p,q) &=&\bar{u}(p) [I_{px} \alpha_x + I_{py} \alpha_y +
(-\mbox{\boldmath $\alpha$} \cdot {\bf p_{\perp}}+ \gamma_0 m) J_p]
\gamma_- u(-q)
 \nonumber \\
&\ &+ \bar{u}(p)[-I_{qx} \alpha_x - I_{qy} \alpha_y +
(\mbox{\boldmath $\alpha$} \cdot {\bf q_{\perp}}+ \gamma_0 m) J_q] 
\gamma_+ u(-q),
\end{eqnarray}
where, letting ${\bf b}$ define the x-axis,
\begin{equation}
I_{px} = {1 \over (2 \pi )^2} \int \exp [i\,  k_x\, b]\,dk_x \,
\int {F_B({\bf k}) \,
F_A({\bf q_{\perp} + p_{\perp} - k})\ k_x\, d k_y \over
-p_+ q_- -({\bf k - p_{\perp}})^2 - m^2 }
\end{equation}
\begin{equation}
I_{py} = {1 \over (2 \pi )^2} \int \exp [i\,  k_x\, b]\,dk_x \,
\int {F_B({\bf k}) \,
F_A({\bf q_{\perp} + p_{\perp} - k})\ k_y\, d k_y \over
-p_+ q_- -({\bf k - p_{\perp}})^2 - m^2 }
\end{equation}
\begin{equation}
I_{qx} = {1 \over (2 \pi )^2} \int \exp [i\,  k_x\, b]\,dk_x \,
\int {F_B({\bf k}) \,
F_A({\bf q_{\perp} + p_{\perp} - k})\ k_x\, d k_y \over
-p_- q_+ -({\bf k - q_{\perp}})^2 - m^2 }
\end{equation}
\begin{equation}
I_{qy} = {1 \over (2 \pi )^2} \int \exp [i\,  k_x\, b]\,dk_x \,
\int {F_B({\bf k}) \,
F_A({\bf q_{\perp} + p_{\perp} - k})\ k_y\, d k_y \over
-p_- q_+ -({\bf k - q_{\perp}})^2 - m^2 }
\end{equation}
\begin{equation}
J_p = {1 \over (2 \pi )^2} \int \exp [i\,  k_x\, b]\,dk_x \,
\int {F_B({\bf k}) \,
F_A({\bf q_{\perp} + p_{\perp} - k})\ d k_y \over
-p_+ q_- -({\bf k - p_{\perp}})^2 - m^2 }
\end{equation}
\begin{equation}
J_q = {1 \over (2 \pi )^2} \int \exp [i\,  k_x\, b]\,dk_x \,
\int {F_B({\bf k}) \,
F_A({\bf q_{\perp} + p_{\perp} - k})\ d k_y \over
-p_- q_+ -({\bf k - q_{\perp}})^2 - m^2 } .
\end{equation}
After squaring, summing over spin states, and taking traces with the aid of
the computer program FORM\cite{form} I obtain the expression for the amplitude
squared
\begin{eqnarray}
\vert M(p,q) \vert^2 &=& p_+ q_- [(m^2 + p_{\perp}^2) \vert J_p \vert^2
+\vert I_{px} \vert^2 +\vert I_{py} \vert^2 - 2 p_x Re(J_p I_{px}^*)
- 2 p_y Re(J_p I_{py}^*)]
\nonumber \\
&+& p_- q_+ [(m^2 + q_{\perp}^2) \vert J_q \vert^2
+\vert I_{qx} \vert^2 +\vert I_{qy} \vert^2 - 2 q_x Re(J_q I_{qx}^*)
- 2 q_y Re(J_q I_{qy}^*)]
\nonumber \\
&+&2 [ (m^2 + p_{\perp}^2)(q_x Re(J_p I_{qx}^*) + q_y Re(J_p I_{qy}^*))
\nonumber \\
&&+(m^2 + q_{\perp}^2) (p_x Re(J_q I_{px}^*) + p_y Re(J_q I_{py}^*))
\nonumber \\
&&+( {\bf p_{\perp} \cdot q_{\perp}} - m^2) (Re(I_{px} I_{qx}^*) + 
Re(I_{py} I_{qy}^*) )
\nonumber \\
&&-(m^2 + p_{\perp}^2)(m^2 + q_{\perp}^2)Re(J_pJ_q^*)   
\nonumber \\
&&-(p_x q_y + p_y q_x) ( Re(I_{px} I_{qy}^*) + Re(I_{py} I_{qx}^*))
\nonumber \\
&&-2 p_x q_x Re(I_{px} I_{qx}^*) -2 p_y q_y Re(I_{py} I_{qy}^*) ] .
\end{eqnarray}.

There is an apparent numerical difficulty in evaluating Equations (9)-(14) due
to the oscillating factor $\exp [i\,  k_x\, b]$.  Of course in the
${\bf b} = 0$ limit this factor is absent, and so we will first investigate
this numerically more tractable case.  The general case of non-zero ${\bf b}$
will then be addressed by a technique involving piecewise analytical
integration.

The derived impact parameter
dependent amplitude squared, $\vert M(p,q) \vert^2$, is not simply the square
of the amplitude for the excitation of a particular (correlated)
electron-positron pair\cite{bg}.   However, as has been discussed in the
literature\cite{read,rmgs,mom,obe,Baur90,Best92,HenckenTB95a,Aste02}
the number operator
for a particular (uncorrelated) electron or positron can be constructed
by integrating over the positron or electron momenta respectively, and likewise
the number operator for total pair production.
In previous articles\cite{ajb3,ajb4} I have also discussed these matters in
more detail.

For the present I will limit myself to calculating the number operator
$<N(b)>$ for total pair production, and from it the total pair production
cross section $\sigma_T$
\begin{equation}
<N(b)> = \int {m^2 d^3 p\, d^3 q  \over (2 \pi)^6 \epsilon_p \epsilon_q }
\vert M(p,q) \vert^2
\end{equation}
Note that $\sigma_T$ corresponds to a peculiar type of inclusive cross section
which we should call the ``number weighted total cross section'',
\begin{equation}
\sigma_T = \int d^2  b <N(b)> = \int d^2 b \sum_{n=1}^{\infty} n P_n(b),
\end{equation}
in contrast to the usual definition of an inclusive total cross section
$\sigma_I$ for pair production,
\begin{equation}
\sigma_I = \int d^2 b \sum_{n=1}^{\infty} P_n(b).
\end{equation}

\section{Eikonal, exact, and perturbative cases}

If one merely regularizes the transverse
integral Eq. (6) at large $\rho$ one obtains\cite{bmc,sw2} 
apart from a trivial phase
\begin{equation}
F({\bf k}) = {4 \pi \alpha Z \over k^{2 - 2 i \alpha Z} }.
\end{equation}
Then all the higher order $Z \alpha$ effects in $M(p,q)$ are contained only
in the phase of the denominator of Eq. (19).  It was noted that as
$Z \rightarrow 0$ this photon propagator leads to the known perturbative
result for $M(p,q)$ if $F({\bf k})$ is modified to
\begin{equation}
F({\bf k}) = {4 \pi \alpha Z \over (k^2+\omega^2 / \gamma^2)^{1-i \alpha Z} }.
\end{equation}
In this approach a lower $k$ cutoff at some $\omega / \gamma$ has
to be put in by hand to obtain dependence on the beam energy and to agree
with the known perturbative result in that limit.

Rather than put in the $\omega^2 / \gamma^2$ cutoff by hand,
my physically motivated ansatz is to apply a spatial cutoff to the
transverse potential $\chi(\mbox{\boldmath $ \rho$})$ (which has been up
to now set to $2 Z \alpha \ln{\rho}$) in order to obtain an expression
consistent with the perturbation theory formula\cite{bs,kai1} in the
ultrarelativistic limit.  In the Weizsacker-Williams or equivalent
photon treatment of
electromagnetic interactions the effect of the potential is cut off at
impact parameter
$b \simeq \gamma / \omega$, where $\gamma$ is the relativistic boost
of the ion producing the photon and $\omega$ is the energy of the photon.
If
\begin{equation}
\chi(\mbox{\boldmath $ \rho$}) = \int_{-\infty}^\infty dz V(\sqrt{z^2+\rho^2})
\end{equation}
and $V(r)$ is cut off in such a physically motivated way, then\cite{lm2}
\begin{equation}
V(r)={-Z \alpha \exp[-r \omega_{A,B} / \gamma] \over r }
\end{equation}
where 
\begin{equation}
\omega_A= {p_+ + q_+ \over 2 };\  \omega_B={p_- + q_- \over 2 }
\end{equation}
with $\omega_A$ the energy of the photon from ion $A$ moving in the positive
$z$ direction and $\omega_B$
the energy of the photon from ion $B$ moving in the negative $z$
direction.  Note that we work in a different gauge than that used to obtain
the original perturbation theory formula, and thus our potential picture is
somewhat different.  The transverse potential will be smoothly cut off at a
distance where the the longitudinal potential delta function approximation is
no longer valid.

\begin{figure}[h]
\begin{center}
\epsfig{file=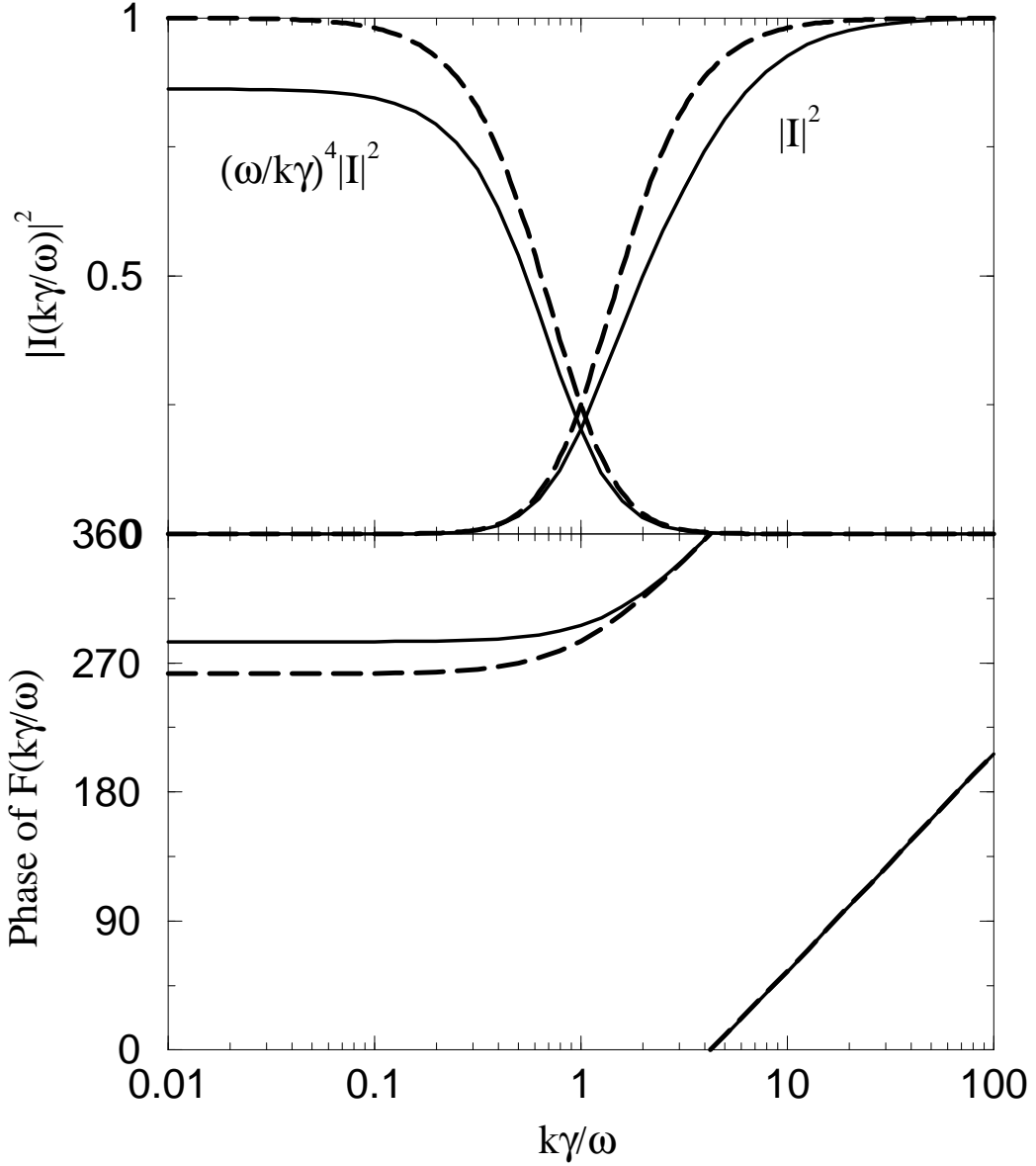,height=16cm}
\end{center}
\caption{Top: Relative magnitude squared of virtual photon source;
solid line exact; long dashed line eikonal or perturbative (see text).
Bottom:  Corresponding phase of virtual photon source; the perturbative
phase (not shown) does not vary with $k \gamma /\omega$.}
\end{figure}
The integral Eq. (21) can be carried out to obtain
\begin{equation}
\chi(\rho)= - 2 Z \alpha K_0(\rho \omega_{A,B} / \gamma),
\end{equation}
and Eq. (6) is modified to
\begin{equation}
F_{A,B}({\bf k}) = 2 \pi \int d \rho \rho J_0(k \rho)
\{\exp [2i Z_{A,B}\alpha K_0(\rho \omega_{A,B} / \gamma)] -1 \} .
\end{equation}
$F_{A}({\bf k})$ and $F_{B}({\bf k})$ are functions of virtual photon
$\omega_A$ and $\omega_B$ respectively.  The modified Bessel function
\begin{equation}
K_0(\rho \omega / \gamma) = -\ln(\rho) -\ln(\omega / 2 \gamma) -\gamma_e
\end{equation}
for $\rho \ll \gamma / \omega $ ($\gamma_e$ is the Euler constant),
and cuts off exponentially at $\rho \approx \gamma / \omega $.  This
is the physical cutoff to the transverse potential.  
One may define $\xi = k \rho$ and rewrite Eq.(25) in terms of a normalized
integral $I_{A,B}(\gamma k / \omega)$
\begin{equation}
F_{A,B}({\bf k}) = {4 \pi i Z_{A,B} \alpha \over k^2} I_{A,B}(\gamma k /
\omega)
\end{equation}
where 
\begin{equation}
I_{A,B}(\gamma k / \omega) =  {1 \over 2 i Z_{A,B} \alpha }
\int d \xi \xi J_0(\xi) \{\exp [2i Z_{A,B}
\alpha K_0(\xi \omega / \gamma k)] - 1 \}.
\end{equation}
Note that $F_{A,B}$ is equal to
$4 \pi Z_{A,B}/k^2$ times a function of $(\gamma k / \omega)$.
The form of the integral, Eq. (6) without the cutoff
(here with added arbitrary phase constants consistent with
$K_0(\xi \omega / \gamma k)$ for small $\xi$,  Eq. (26))
was solved in closed form\cite{bmc,sw2}
\begin{equation}
I^E_{A,B}(\gamma k / \omega) =
-i ({\exp[\gamma_e] \omega \over \gamma k})^{-2 i \alpha Z} 
{\Gamma(-i \alpha Z) \over \Gamma(i \alpha Z)}
{ 1 \over (1 + \omega^2 / k^2 \gamma^2)^{1 -i \alpha Z}}.
\end{equation}
I will refer to this form as the eikonal.
In the limit as $Z\to0$ of $I^E_{A,B}(\gamma k / \omega)$ goes to
\begin{equation}
I^0_{A,B}(\gamma k / \omega) = { -i \over 1 + \omega^2 / k^2 \gamma^2},
\end{equation}
leading to the familiar perturbation theory form 
\begin{equation}
F^0_{A,B}({\bf k}) = {4 \pi i Z_{A,B} \alpha \over k^2 + \omega^2 / \gamma^2}.
\end{equation}

The top panel of Figure 1 displays the results of numerical calculation
of $\vert I(k \gamma / \omega) \vert^2$ for $Z = 79$ and in
the perturbative limit.  The curves
that go to zero on the left are $\vert I(k \gamma / \omega) \vert^2$.
The curves that go to zero on the right are 
$\vert I(k \gamma / \omega) \vert^2$ multiplied by $(\omega / k \gamma)^4$
to exhibit the reduction at high $Z$ on the left of the curve.
$(\omega / k \gamma)^4 \vert I(k \gamma / \omega) \vert^2$ goes to one
as $k \gamma / \omega$ goes to zero in the
perturbative case; it goes to a reduced constant value as $k \gamma / \omega$
goes to zero for $Z=79$.  The bottom panel exhibits the phase of the full
expression $F({\bf k})$.  The perturbative phase (not shown) is constant.
Both exact and eikonal phases are identical at high $k \gamma / \omega$
and diverge by 22 degrees for this $Z=79$ case as they approach constants at
low $k \gamma / \omega$.

\section{The ${\bf b} = 0$ Limit}

Since the present calculations, based on the ultrarelativistic limit, employ a
somewhat different approach than previous perturbative calculations, but should
agree with them in the perturbative limit, I have deemed it prudent to do
some comparison of numerical results.   As a first step I computed the
${\bf b} = 0$ limit, both as a test of the method and general structure of
the computer code as well as for physical interest.  Figure 2
shows a pertubation theory calculation of the energy spectrum of produced
positrons at ${\bf b} = 0$ compared with a previous perturbation theory
calculation of Hencken, Trautmann, and Baur\cite{kai2}.  The same dipole
form factor has been utilized for the ions with $\Lambda = 83$ MeV, and the
curves have been divided by $(Z \alpha)^4$.
The heavy ions are
from Au + Au at RHIC, and the energy of the electrons has been integrated over.
Except at the lowest
energies the agreement is good, and the integrated probability of the present
calculation $P^0(0) = 1.64$ is in good agreement with the previous
calculation $P^0(0) = 1.6$.  The integrated numbers have not been divided
by $(Z \alpha)^4$ as the displayed curves were.

\begin{figure}[h]
\begin{center}
\epsfig{file=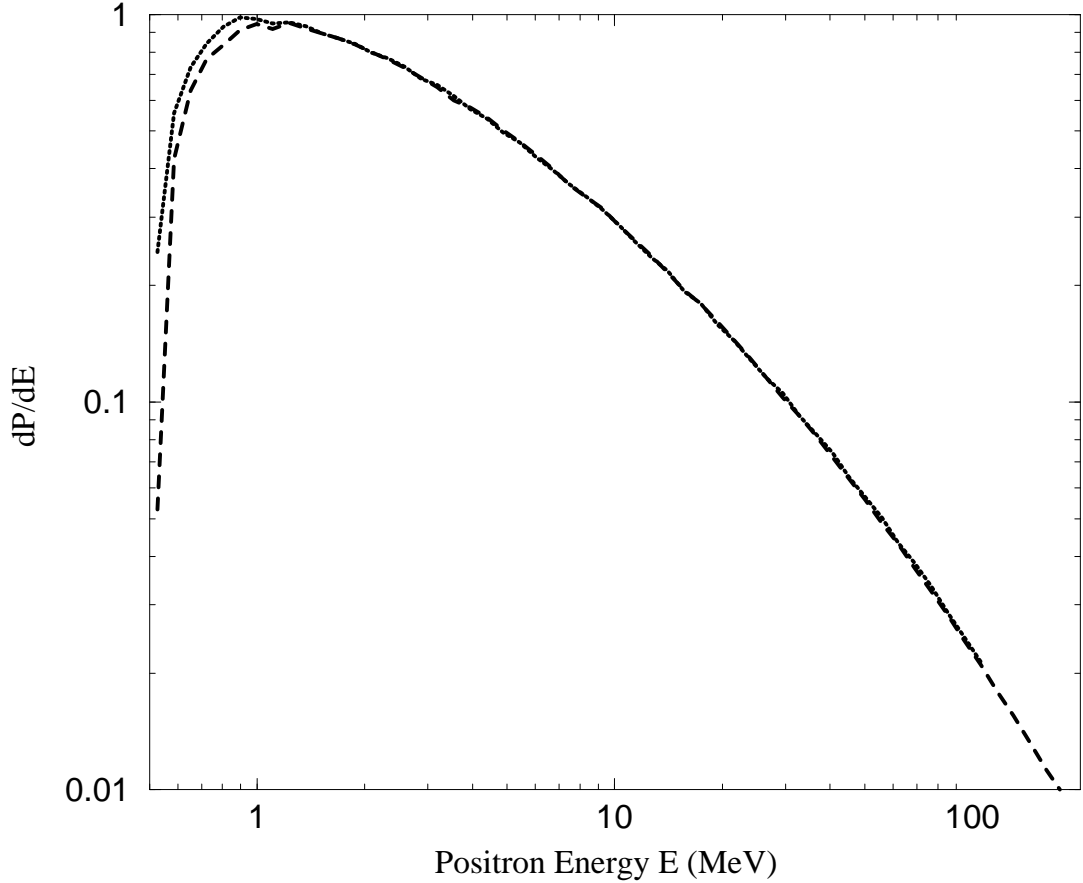,height=12cm}
\end{center}
\caption{Total probability as a function of energy of the positron.  The
curves have been divided by $(Z \alpha)^4$. Short dashes: present perturbation
theory; Dotted line: previous perturbation theory calculation\cite{kai2}}
\end{figure}
Figure 3 shows the effect of higher order contributions.  My perturbation
result of Figure 2 is repeated for reference.  The solid line is the result
of the exact calculation.  It is clearly reduced from perturbation theory
for the lowest to highest values of the energy range; the exact energy 
integrated $P(0) = .94 = .57 P^0(0)$.  Only slightly larger in magnitude
is the result of an eikonal calculation where the energy integrated
$P^E(0) = 1.03 = .63 P^0(0)$.
\begin{figure}[h]
\begin{center}
\epsfig{file=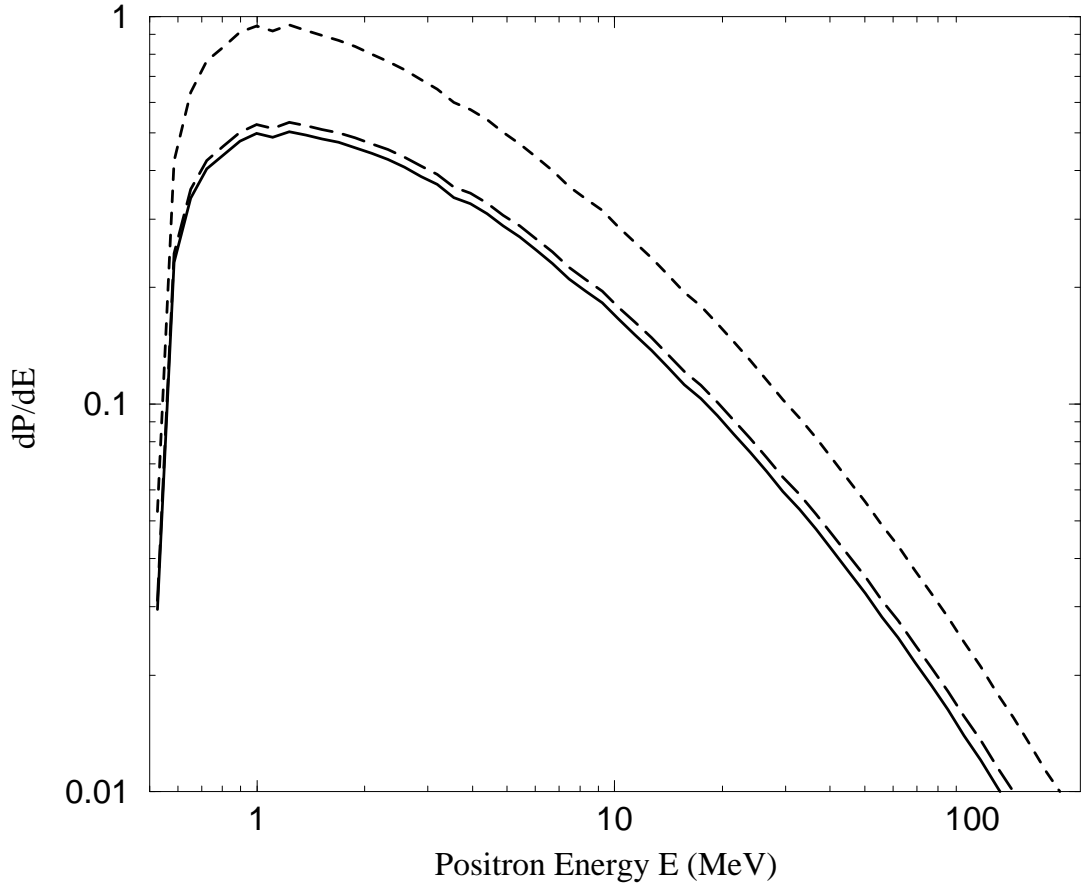,height=12cm}
\end{center}
\caption{Probabilities as in Fig.2. Short dashes: present perturbation theory;
Solid line: exact); Long dashes: eikonal}
\end{figure}

The overall reduction from perturbation theory is almost as large
in the eikonal case as it is in the exact calculation, and the two curves
in Figure 3 do not diverge much at any energy value.  This would suggest that
similar effective physics is involved in the two calculations.
Recall from Figure 1 the reduction in magnitude of the virtual photon
source for the exact calculation as compared to the eikonal (which is
identical in magnitude to perturbation theory) for $k \gamma / \omega < 50$.
For $k >\  100\ \omega / \gamma $ the
magnitude of $F({\bf k})$ goes over into the eikonal or perturbative result,
while the phase of $F({\bf k})$ is identical to the eikonal.  It is in this
large $k$ region, where $F({\bf k})$ becomes identical the eikonal
in both magnitude and phase, that most of the contribution for ${\bf b} = 0$
comes.  Test calculations show that 80\% of the contribution to the total
probability at ${\bf b} = 0$ comes from $k >\ 100\ \omega / \gamma $ and that
less than a tenth of a percent comes from $k <\ 10\ \omega / \gamma $.
It is not the region of reduced magnitude that dominates for the exact
case ${\bf b} = 0$, but the region of rotating phase, and that
rotating phase is what reduces
the cross section from perturbation theory.  The eikonal
$F({\bf k})$ is a fair approximation to the exact at ${\bf b} = 0$, but as we
will see in Section V this is not true for larger ${\bf b}$.

It is worth emphasizing that calculations labeled perturbative, exact, and
eikonal
differ only in the expressions used for $F_{A,B}({\bf k})$.  The analytical
expression Eq. (31) is used for perturbative calculations, and 
Eq. (27) and Eq. (29) for the eikonal calculations.  The exact calculations
makes use the expression Eq. (27) and Eq. (28), which must be evaluated
numerically, but only once for each $Z_{A,B}$ of interest.

\section{${\bf b}$ Dependent Probabilities and the total $e^+ e^-$ cross
section: ${\bf Au + Au}$ at RHIC}

With the addition of impact parameter dependence, we are faced with a nine
dimensional integral for the total cross section as compared to the seven
dimensional integral in the previous section, or the seven dimensional
integral in the representation of cross sections without impact
parameter\cite{ajb4}.
Although the usual method of evaluation, e.g. in perturbation
theory, is via Monte Carlo, I have chosen to do the multi dimensional
integral directly on meshes uniform on a logarithmic scale in each radial
momentum dimension and on a logarithmic scale in impact parameter.  It was
computationally tedious but possible to carry out the calculations to some
rough but significant accuracy without using Monte
Carlo because the integrands are very smooth and smoothly goes to zero at both
high end and low end of the momentum ranges and impact parameter range.

Apart from the fact that impact parameter dependence adds two more dimensions
to calculations (the magnitude of ${\bf b}$ and the angle of the outgoing pair
relative to ${\bf b}$), evaluation of the square of the pair production
amplitude, Eq. (15), is
considerably more difficult for non-zero impact parameter $b$ due to
the factor $\exp [i\,  k_x\, b]$ in Eq. (9-14).  For increasing values of
${\bf b}$ the integral over $k_x$ oscillates rapidly and presents a challenge
to numerical integration.  Furthermore, while the oscillation is on a linear
scale in $k_x$, experience has shown that the natural scale for integration
over momenta in this electromagnetic problem is logarithmic; we want to choose
numerical mesh points on a logarithmic scale.  I have chosen to work in
Cartesian coordinates
because after a numerical integration over the comparative smooth variation
in $k_y$ I am able to make use of a relatively simple piecewise analytical
method to integrate over $k_x$.  I do the integration over $k_x$ in the
manner described in Appendix A.

\begin{figure}[h]
\begin{center}
\epsfig{file=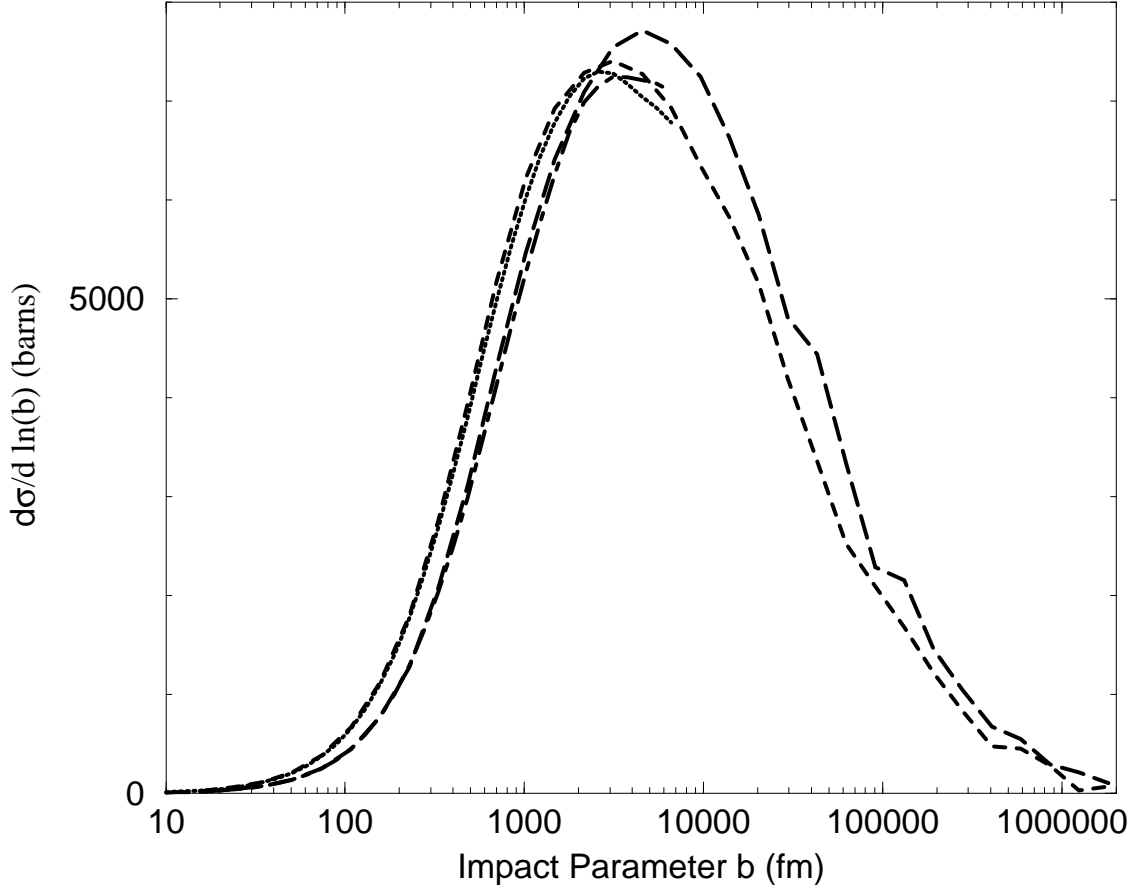,height=12cm}
\end{center}
\caption{Impact parameter dependence of contribution to total cross section.
Dashed  line, perturbation theory; long dashed line eikonal.  Comparable
results are shown derived from the calculations of Hencken, Trautmann, and
Baur\cite{kai}: Dotted line, perturbation theory; Dot dashed line, eikonal.}
\end{figure}
\begin{figure}[h]
\begin{center}
\epsfig{file=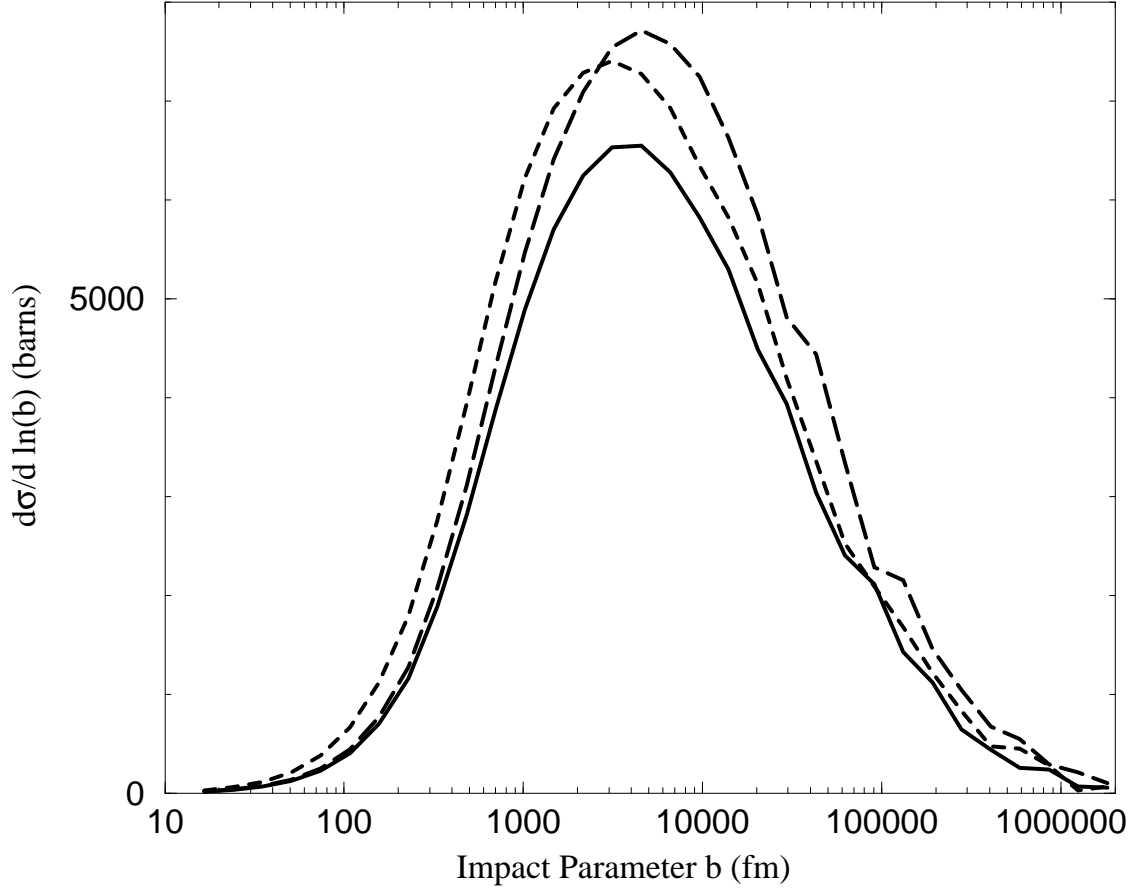,height=12cm}
\end{center}
\caption{Impact parameter dependence of contribution to total cross section.
Dashed  line, perturbation theory; long dashed line eikonal; solid line,
exact.}
\end{figure}
\begin{figure}[h]
\begin{center}
\epsfig{file=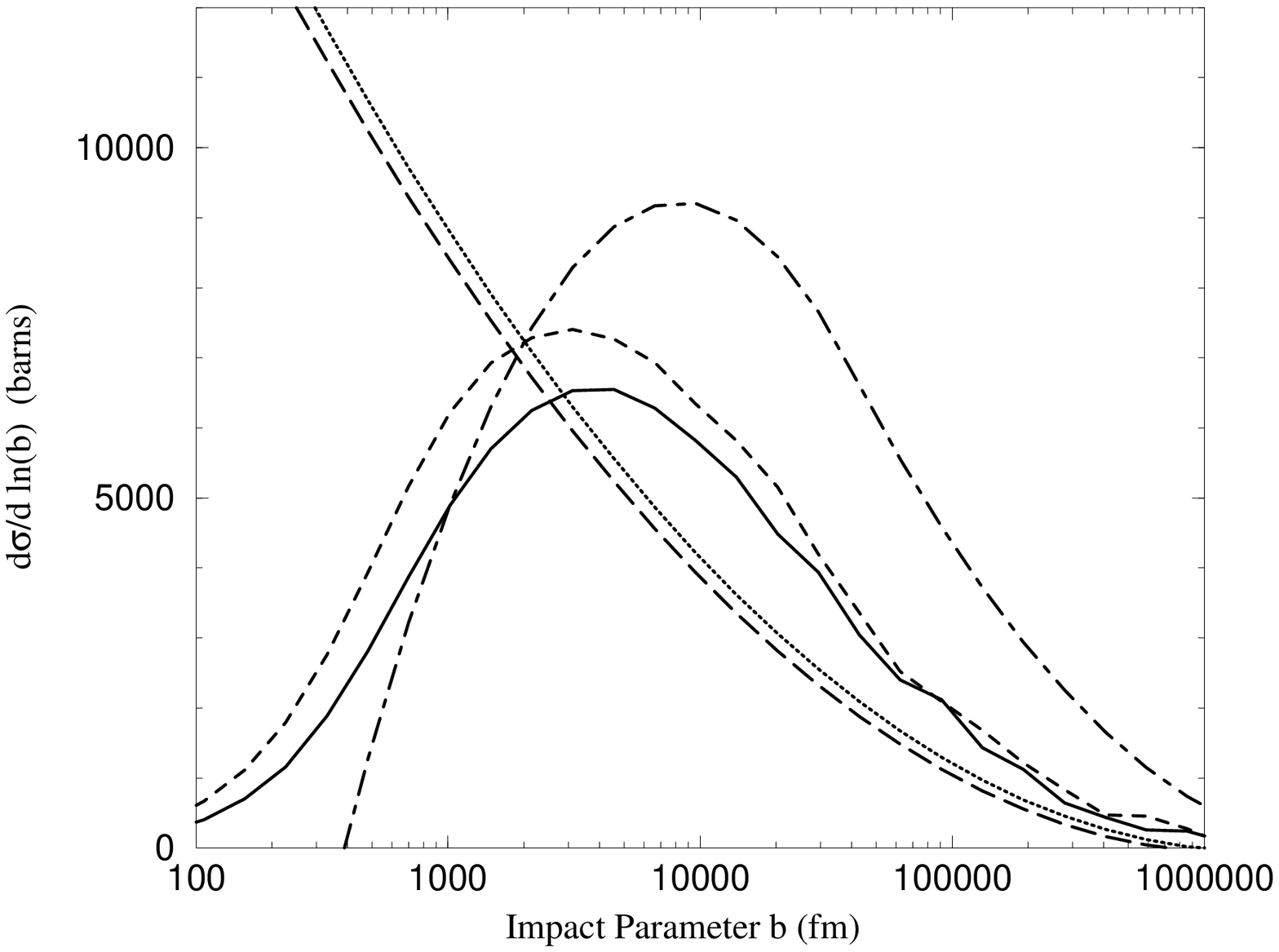,height=12cm}
\end{center}
\caption{Impact parameter dependence of contribution to total cross section.
Present calculation: Dashed  line, perturbation theory; Solid line,
exact.  Comparable results are shown derived from the calculations of
Bertulani and Baur\cite{bb}: Dotted line, perturbation theory; Long dashed
line, Coulomb corrected.  Comparable perturbative results derived from the
formula of Lee, Milstein, and Serbo\cite{lms} are represented by the
dot-dashed line.}
\end{figure}

Calculations were again carried out for the RHIC case of Au + Au at
$\gamma = 100$.
For the perturbation theory and eikonal probabilities I was able to do a
rough check of my computer code by comparing with another set of
published calculations\cite{kai}.  Figure 4 shows the comparison of
cross section contributions derived from the
present calculations with those derived from the ${\bf b}$ dependent
probabilities of Hencken, Trautmann, and Baur.  The logarithmic derivative
was chosen so that contributions to the integrated cross section go as the
area under the curves on this natural log scale plot.
No form factor was utilized in
either calculation as its effect should be small for total cross section
probabilities at these impact parameters.  Agreement is reasonably good,
considering the large grids used in the present calculations.  Note that
that for both sets of calculations the eikonal solution is less that
perturbation theory for low impact parameters, and then crosses over to be
greater than perturbation theory at about $b = 3000$ fm and larger.
For small impact parameters there is fair agreement, but there is an incipient
divergence between the curves for the present calculations
and those of Hencken, Trautmann, and Baur where the latter cut off.
However, one of the authors had warned me not to trust their results far out 
in b too much because the Fourier transform used was
not as good there\cite{kaip}.

In Figure 5 one can observe the comparison of the exact calculation with
perturbation theory and the eikonal.  For all impact parameters the
probability of pair production is
smaller in the exact calculation than it is in perturbation theory.
The small impact parameter agreement between the exact calculation and
the eikonal extends the mechanism discussed in the ${\bf b} = 0$ case, the
reduction from pertubation theory due to the similar rotating phase of $F(k)$
for dominant large values of $\bf{k}$ seen in Fig. 1.  For larger impact
parameters providing the dominant contribution to the cross section,
a significant contribution comes in the region
$  k <\ 10\ \omega / \gamma $ 
where the magnitude of  $F({\bf k})$ is reduced from perturbation theory
and the eikonal.  This demonstrates how a correct physical cutoff process
preserved the Coulomb corrections at high ${\bf b}$ and thus leads to
Coulomb corrections in the total cross section.

The calculations can be integrated over impact parameter to compare with
my previous calculations\cite{ajb3} done not in the impact parameter
representation.  That the present integrated perturbation theory calculation
of 34.6 kbn (kilobarns) is in such good agreement with my previous calculation
of 34.6 kbn must be considered fortuitous, given the relatively crude nature
of the impact parameter calculation.  The present integrated exact calculation
of 29.4 kbn is in fair agreement with my previous 28.6 kbn.  The integrated
eikonal calculation of 35.5 kbn should in principle be identical to the
34.6 kbn perturbation theory calculation, but like the present exact
calculation is slightly larger than
my previous published non impact parameter calculation.

At this point it is interesting to also compare the present
results with two previous approximate analytical expressions for impact
parameter dependence.  In their classic Physics Reports article, Bertulani
and Baur\cite{bb} presented an impact
parameter formula that included a Coulomb correction.
Figure 6 shows a comparison of the cross section contribution
derived from their probability formula with the present results.
Bertulani and Baur integrated over their stated range of validity,
from $1/m_e$ to $\sim \gamma^2/m_e$, to obtain a Landau-Lifshitz
perturbative result with a
Coulomb correction.  For symmetric heavy ions, the Coulomb correction that
they obtained was of the same form as the result later obtained by
Serbo et al.\cite{serb} and by Lee and Milstein\cite{lm1,lm2}.  However,
since Bertulani and Baur had postulated averaging the result of a
Coulomb correction for the target with one for the projectile,
they obtained half of the more correct later result based on adding the
lowest order Coulomb correction for both target and projectile.
The overall shapes of the Bertulani and Baur results differ
from the present results, but their Coulomb correction reduces the
probability throughout the impact parameter range.

Shortly after obtaining their above mentioned Coulomb correction results for
the total cross section\cite{serb,lm1,lm2} Lee, Milstein,and Serbo presented
an analytical form for the the impact parameter dependence\cite{lms} that
differed substantially from that of Bertulani and Baur, especially
in the intermediate range from $1/m_e$ to $\gamma/m_e$.  Figure 6 also displays
the impact parameter contribution to the cross section obtained from their
pertubative expression for the probability.  The shape of the curve seems
qualitatively more in agreement with present results.  As in the case of
Bertulani and Baur, the treatment is considered valid for values of
impact parameter larger than $1/m_e$ (386 fm), and in both cases
integration of the contribution from their expression is
in agreement with the leading $\ln^3(\gamma^2)$ term of Landau and
Lifshitz\cite{ll}.  Lee, Milstein, and Serbo
did not present a specific form for the impact parameter dependence of
the Coulomb correction, but as noted above, their previous work indicates
a negative Coulomb correction on average
twice that of Bertulani and Baur over the indicated parameter range.

\section{Summary and Discussion}

Calculated exact total probabilities for heavy ion $e^+ e^-$ pair
production exhibit a reduction from the probabilities calculated in
perturbation theory throughout the full range of impact parameters.

In principle the impact parameter approach to calculating exact $e^+ e^-$ pair
production probabilities is suited for combining with
an impact parameter dependent mutual Coulomb dissociation
calculation\cite{bcw} to be able to compare with e.g. data such as was obtained
with the Solenoidal Tracker at RHIC (STAR) setup\cite{star04}.  In this paper
I have presented calculations only for the total $e^+ e^-$ pair production
probabilities.  Still to be done is a sufficiently accurate
calculation of the high transverse momentum slice of data seen by STAR
to be combined with a Coulomb dissociation calculation for the zero degree
calorimeter acceptance.  Also since the present approach is strictly
speaking valid only when either the positrons or electons have been
integrated over, and in the STAR case both electron and positron are
constrained to be in the high momentum slice, the present approach to
$e^+ e^-$ production is not exactly valid for the STAR case.  At present
the best one can do is observe that the present method is valid for both
uncorrelated positrons and electrons of all momenta, and ignore the
correlations.  The effect of correlations averages to zero, but some
estimate of individual magnitudes would be useful.

\section{Acknowledgments}
I would like to thank Francois Gelis for useful discussions.  I would like
to thank Kai Hencken for providing numerical files of the results of
References \cite{kai2} and \cite{kai} and for useful discussions.

This manuscript has been authored
under Contract No. DE-AC02-98CH10886 with the U. S. Department of Energy. 

\appendix
\section{Piecewise analytical Fourier integration}
Consider the integral over $k_x$ (I will drop the subscript $x$ for now).
It contains a smoothly varying part which I will call $f(k)$ times the
rapidly varying coefficient $\exp [i\,  k\, b]$
\begin{equation}
I = \int_{-\infty}^{\infty} f(k) \exp [i\,  k\, b] dk.
\end{equation}
One can use integration by parts to transform $I$ to
\begin{equation}
I = \int_{-\infty}^{\infty}{\bigl(1 - \exp [i\,  k\, b]\bigr) \over i b}
{df(k)\over dk} dk,
\end{equation}
where the term $1$ is inserted for numerical convenience.
One now rewrites $I$ in a form suggesting piecewise analytical integration
from mesh point to mesh point 
\begin{equation}
I = \sum_{k_{i{\rm min}}}^{k_{i{\rm max}}} \int_{k_i}^{k_{i+1}}
{\bigl(1 - \exp [i\,  k\, b]\bigr) \over i b} {df(k)\over dk} dk.
\end{equation}
Taking the lowest order (constant) approximation to the derivative over the
interval one has
\begin{eqnarray}
I &=& \sum_{k_{i{\rm min}}}^{k_{i{\rm max}}} \biggl({f(k_{i+1})-f(k_i)
\over k_{i+1} - k_i}\biggr)
\int_{k_i}^{k_{i+1}} {\bigl(1 - \exp [i\,  k\, b]\bigr) \over i b} dk
\nonumber \\
&=& \sum_{k_{i{\rm min}}}^{k_{i{\rm max}}} \biggl({f(k_{i+1})-f(k_i)
\over k_{i+1} - k_i}
\biggr)
\biggl( { k_{i+1} - k_i \over i b}\,+\,
{\exp [i\,  k_{i+1}\, b] - \exp [i\,  k_i\, b] \over b^2} \biggr). 
\end{eqnarray}
This is the lowest order expression for piecewise analytical integration.

One can obtain the piecewise analytical expression good to the next leading
order by taking three point Lagrange interpolation for $f(k)$ between $k_i$
and $k_{i+1}$ and then taking the 
derivative.  One obtains
\begin{eqnarray}
I&=&\sum_{k_{i{\rm min}}}^{k_{i{\rm max}}} {1 \over i b} \biggl(A\, 
(k_{i+1} - k_i)
\,+\,{B \over 2}(k_{i+1}^2 - k_i^2) \nonumber \\
&-&{\exp [i\,  k_{i+1}\, b]\,\Bigl({A+B\,k_{i+1} \over i b}}
+{B \over b^2}\Bigr)\,+\,{\exp [i\,  k_i\, b]\,\Bigl({A+B\,k_i \over i b}}
+{B \over b^2}\Bigr)
\,\biggr),
\end{eqnarray}
where
\begin{equation}
A=-\,\biggl({f(k_i)\,(k_{i+1} + k_{i+2}) \over (k_i - k_{i+1})\,
(k_i - k_{i+2})}
+{f(k_{i+1})\,(k_i + k_{i+2}) \over (k_{i+1} - k_i)\,(k_{i+1} - k_{i+2})}
+{f(k_{i+2})\,(k_i + k_{i+1}) \over (k_{i+2} - k_i)\,(k_{i+2} - k_{i+1})}
\biggr),
\end{equation}
and
\begin{equation}
B=2\,\biggl({f(k_i) \over (k_i - k_{i+1})\,(k_i - k_{i+2})}
+{f(k_{i+1}) \over (k_{i+1} - k_i)\,(k_{i+1} - k_{i+2})}
+{f(k_{i+2}) \over (k_{i+2} - k_i)\,(k_{i+2} - k_{i+1})}
\biggr).
\end{equation}

The expression Eq. (A5-A7) exhibits numerical difficulties for small impact
parameters.  Therefore, for impact parameters in the range below the
electron Compton wavelength (about 386 fm.) I have utilized Eq. (A4), and
I have utilized Eq. (A5-A7) for higher impact parameter values.


\begin{thebibliography}{99}

\bibitem {pes} Michael E. Peskin and Daniel V. Schroeder, {\it An Introduction
to Quantum Field Theory}, Sec. 4.5 (1995)
\bibitem {bcw} A. J. Baltz, C. Chasman, and S. N. White, Nucl. Instrum.
Methods Phys. Res., Sect. A 417, 1 (1998).
\bibitem {brbw} A. J. Baltz, M. J. Rhoades-Brown, and J. Weneser,
Phys. Rev. {\bf A 44}, 5569 (1991).
\bibitem {ajb1} A. J. Baltz, Phys. Rev. {\bf A 52}, 4970 (1995).
\bibitem {ajb2} A. J. Baltz, Phys. Rev. Lett. {\bf 78}, 1231 (1997).
\bibitem {sw1} B. Segev and J. C. Wells, Phys. Rev. {\bf A 57}, 1849 (1998).
\bibitem {bmc} Anthony J. Baltz and Larry McLerran, Phys. Rev. {\bf C 58},
1679 (1998).
\bibitem {sw2} B. Segev and J. C. Wells, Phys. Rev. {\bf C 59}, 2753 (1999).
\bibitem {serb} D. Yu. Ivanov, A. Schiller, and V. G. Serbo, Phys. Lett. B
{\bf 454}, 155 (1999).
\bibitem {lm1} R. N. Lee and A. I. Milstein, Phys. Rev. {\bf A 61}, 032103
(2000).
\bibitem {lm2} R. N. Lee and A. I. Milstein, Phys. Rev. {\bf A 64}, 032106
(2001).
\bibitem {ll} L. D. Landau and E. M. Lifshitz, Phys. Z. Sowjetunion 6,
244(1934).
\bibitem {bem} H. A. Bethe and L. C. Maximon, Phys. Rev. {\bf 93}, 768 (1954);
Handel Davies, H. A. Bethe and L. C. Maximon, Phys. Rev. {\bf 93}, 788 (1954).
\bibitem {bb} Carlos A. Bertulani and Gerhard Baur,
Physics Reports {\bf 163}, 299 (1988).
\bibitem {ajb3} A. J. Baltz, Phys. Rev. {\bf C 68}, 034906 (2003).
\bibitem {ajb4} A. J. Baltz, Phys. Rev. {\bf C 71}, 024901 (2005).
\bibitem {form} J. A. M. Vermaseren, arXiv:math-ph/0010025 (2000).
\bibitem {bg} A. J. Baltz, F. Gelis, L. McLerran, and A. Peshier,
Nucl. Phys. {\bf A 695}, 395 (2001).
\bibitem {read} J. F. Reading, Phys. Rev. {\bf A 8}, 3262 (1973).
\bibitem {rmgs} J. Reinhardt, B. M\"uller, W. Greiner, and G. Soff,
Phys. Rev. Lett. {\bf 43}, 1307 (1979).
\bibitem {mom} Klaus Rumrich, Klaus Momberger, Gerhard Soff, Walter Greiner,
Norbert Gr\"un, and Werner~Scheid, Phys. Rev. Lett. {\bf 66}, 2613 (1991).
\bibitem {obe} J. C. Wells, V. E. Oberacker, A. S. Umar, C. Bottcher,
M. R. Strayer, J. S. Wu, G. Plunien, Phys. Rev. {\bf A 45}, 6296, (1992).
\bibitem {Baur90}G. Baur, Phys. Rev.~A {\bf 42},  5736  (1990).
\bibitem {Best92}C. Best, W. Greiner, and G. Soff, Phys. Rev.~A {\bf 46},
261  (1992).
\bibitem {HenckenTB95a}
K. Hencken, D. Trautmann, and G. Baur, Phys. Rev.~A {\bf 51},  998  (1995).
\bibitem {Aste02} A. Aste, G. Baur, K. Hencken, D. Trautmann, and G. Scharf,
Eur. Phys. J. C {\bf 23}, 545 (2002).
\bibitem {bs} C. Bottcher and M. R. Strayer, Phys. Rev. {\bf D 39},
1330 (1989).
\bibitem {kai1} Kai Hencken, Dirk Trautmann, and Gerhard Baur, 
Phys. Rev. {\bf A 51}, 1874 (1995).
\bibitem {kai} Kai Hencken, Dirk Trautmann, and Gerhard Baur,
Phys. Rev. {\bf C 59}, 841 (1999).
\bibitem {kai2} Kai Hencken, Dirk Trautmann, and Gerhard Baur, 
Phys. Rev. {\bf A 49}, 1584 (1994).
\bibitem {kaip}  Kai Hencken, private communication.
\bibitem {lms}  R. N. Lee, A. I. Milstein, and V. G. Serbo,
Phys. Rev. {\bf A 65}, 022102 (2002).
\bibitem{star04} STAR Collaboration, J. Adams {\it et al.},
Phys. Rev.~C {\bf 70}, 031902(R) (2004).
\end{thebibliography}
\end{document}